\documentclass[fluids,article,accept,oneeauthor,pdftex]{Definitions/mdpi} 

\firstpage{1} 
\pubvolume{1}
\issuenum{1}
\articlenumber{0}
\pubyear{2021}
\copyrightyear{2021}
\datereceived{} 
\dateaccepted{} 
\datepublished{} 
\hreflink{https://doi.org/} 
\pdfoutput=1

\usepackage{bm}
\usepackage[normalem]{ulem}

\Title{Fluid Dynamics in Curvilinear Coordinates without Fictitious Forces}

\TitleCitation{Fluid Dynamics in Curvilinear Coordinates without Fictitious Forces}

\Author{Christian Y. Cardall $^{1,\dagger}$\orcidA{}}

\AuthorNames{Christian Y. Cardall}

\AuthorCitation{Cardall, C. Y.}

\address{%
$^{1}$ \quad Physics Division, Oak Ridge National Laboratory, Oak Ridge, TN 37831-6354, USA; cardallcy@ornl.gov}

\corres{cardallcy@ornl.gov}

\firstnote{This manuscript has been authored by UT-Battelle, LLC, under contract DE-AC05-00OR22725 with the US Department of Energy (DOE). The US government retains and the publisher, by accepting the article for publication, acknowledges that the US government retains a nonexclusive, paid-up, irrevocable, worldwide license to publish or reproduce the published form of this manuscript, or allow others to do so, for US government purposes. DOE will provide public access to these results of federally sponsored research in accordance with the DOE Public Access Plan (\texttt{http://energy.gov/downloads/doe-public-access-plan}).}

\abstract{Use of curvilinear coordinates is sometimes indicated by the inherent geometry of a fluid dynamics problem, but this introduces fictitious forces into the momentum equations that spoil strict conservative form.
If one is willing to work in three dimensions, these fictitious forces can be eliminated by solving for rectangular (Cartesian) momentum components on a curvilinear mesh.
A thoroughly geometric approach to fluid dynamics on spacetime demonstrates this transparently, while also giving insight into a greater unity of the relativistic and nonrelativistic cases than is usually appreciated.}


\begin{document}

\section{Introduction}
\label{sec:Introduction}

A fluid, like any material continuum, is defined and governed by tensor field equations on spacetime, a four-dimensional differentiable manifold \cite{Landau1987Fluid-Mechanics,Ferrarese2008Introduction-to}.
The fluid's existence and kinematics are defined by the vanishing 4-divergence of a matter flux vector $\bm{N}$,
\begin{equation}
\bm{\nabla} \cdot \bm{N} = 0,
\label{eq:MatterDivergence_4}
\end{equation}
expressing conservation of the basic `stuff' underlying the fluid (typically mass, or better, baryon number).
The dynamics of the fluid is governed by Newton's second law and the first law of thermodynamics as applied to individual fluid elements. 
Given the 4-velocity field $\bm{w}$ defining the worldlines of fiducial observers, and in the absence of external forces and heating/cooling, these laws governing momentum and energy evolution can be combined in terms of the conservation law
\begin{equation}
\bm{\nabla} \cdot \bm{S} = 0
\label{eq:EnergyMomentumDivergence_4}
\end{equation}
for the relative energy-momentum flux, a $(1,1)$ tensor $\bm{S}$ \cite{Cardall2019Minkowski-and-G,Cardall2020Combining-3-Mom}.
The label ``relative energy-momentum flux'' denotes the fact that the energy content of $\bm{S}$ includes only internal energy and bulk motion relative to fiducial observers $\bm{w}$, allowing the same Equation~(\ref{eq:EnergyMomentumDivergence_4}) on spacetime to apply, regardless of whether causality is governed by absolute time (the ``non-relativistic'' world of Galilei and Newton) or by light cones (the ``relativistic'' world of Einstein and Minkowski) \cite{Cardall2019Minkowski-and-G,Cardall2020Combining-3-Mom}.
Even though there is no spacetime metric in the Galilei/Newton case, the 4-divergence still exists thanks to the presence of a 4-volume form (Levi-Civita tensor) $\bm{\mathcal{E}}$, through which 
\begin{equation}
\left( \bm{\nabla} \cdot \bm{A} \right) \bm{\mathcal{E}} = \bm{\mathrm{d}} \left( \bm{A} \cdot \bm{\mathcal{E}}  \right)
\end{equation}
defines the divergence of a vector field $\bm{A}$; the right-hand side is the exterior derivative of the interior product of $\bm{A}$ with $\bm{\mathcal{E}}$. 

To address an initial value problem, foliate spacetime into spacelike slices.
For present purposes let spacetime be flat, a 4-dimensional affine space, either Galilei/Newton spacetime $\mathbb{G}$ equipped with absolute time \cite{Weyl1922Space---Time---,Cartan1923Sur-les-variete,Cartan1924Sur-les-variete,Cartan1986On-manifolds-wi,Toupin1957World-invariant,Truesdell1960The-Classical-F,Trautman1965Foundations-and,Trautman1966Comparison-of-N,Kunzle1972Galilei-and-Lor} or Minkowski spacetime $\mathbb{M}$ equipped with a Lorentz metric \cite{Gourgoulhon2013Special-Relativ}.
In these cases we can choose the fiducial observers to be inertial observers, with worldlines given by the straight coordinate curves of a global time coordinate $t$, and 4-velocities $\bm{w} = \bm{\partial}_t$ given by the natural basis vectors $\bm{\partial}_t$ tangent to this congruence of straight worldlines.
On $\mathbb{G}$ the coordinate $t$ is the absolute time, and the spacelike slices are already given as its level surfaces.
On $\mathbb{M}$ the coordinate $t$ is the proper time of the fiducial (and here, inertial) observers; and by virtue of the Lorentz metric, choose to specify that the spacelike slices be orthogonal to $\bm{w}$, and thus level surfaces of $t$ in this case as well.
On both $\mathbb{G}$ and $\mathbb{M}$ these spacelike slices $\left( \mathbb{S}_t \right)$ are three-dimensional affine hyperplanes equipped with a Euclidean 3-metric $\bm{\gamma}$.
And in both cases the 4-volume form reads
\begin{equation}
\bm{\mathcal{E}} = \bm{\mathrm{d}}t \wedge \bm{\varepsilon},
\end{equation}
in which
\begin{equation}
\bm{\varepsilon} = \bm{\mathrm{d}}x \wedge \bm{\mathrm{d}}y \wedge \bm{\mathrm{d}}z 
	= \sqrt{\gamma} \; \bm{\mathrm{d}}u \wedge \bm{\mathrm{d}}v \wedge \bm{\mathrm{d}}w
\end{equation}
is the 3-volume form on the slices $\left( \mathbb{S}_t \right)$, expressed in terms of rectangular coordinates $\left( x^{\bar\imath} \right) = \left( x,y,z \right)$ or curvilinear coordinates $\left( u^i \right) = \left( u,v,w \right)$, with $\gamma$ being the determinant of the matrix of curvilinear metric components.
(Throughout this paper, latin indices denote coordinates, and also components of tensors, on the $\left( \mathbb{S}_t \right)$. 
Indices with an overbar are associated with rectangular coordinates, while unadorned indices denote curvilinear coordinates.
Writing  the $\left( x^{\bar\imath} \right)$ as functions of the $\left( u^i \right)$, the elements of the matrix of curvilinear metric components $\left[ \gamma_{ij} \right]$ are read off the line element $dx^2 + dy^2 + dz^2 = \gamma_{a b} \, du^a \, du^b$. 
 Note our use throughout of the Einstein summation convention; for readability, repeated summation indices $a, b, c, \dots$ are taken from near the beginning of the alphabet, while $i, j, k, \dots$ denote free indices.)
 
This foliation of spacetime---here, into affine hyperplanes of $\mathbb{G}$ and $\mathbb{M}$---effects a $3+1$ decomposition of Equations~(\ref{eq:MatterDivergence_4}) and (\ref{eq:EnergyMomentumDivergence_4}) in terms of matter, momentum, and energy densities $N$, $\bm{\Pi}$, and $E$ respectively, and corresponding spatial fluxes $\bm{F}_N$, $\bm{F}_{\bm{\Pi}}$, and $\bm{F}_E$ tangent to $\mathbb{S}_t$, measured by the fiducial observers:
\begin{eqnarray}
\frac{\partial N}{\partial t} + \bm{\mathrm{D}} \cdot {\bm F}_N &=& 0, 
\label{eq:Matter_31} \\
\frac{\partial \bm{\Pi}}{\partial t} + \bm{\mathrm{D}} \cdot {\bm F}_{\bm{\Pi}} &=& 0, 
\label{eq:Momentum_31} \\
\frac{\partial E}{\partial t} + \bm{\mathrm{D}} \cdot {\bm F}_E &=& 0, 
\label{eq:Energy_31} 
\end{eqnarray}
in which the 3-divergences $\bm{\mathrm{D}} \cdot \bm{F}$ are on $\mathbb{S}_t$, and $\sqrt{\gamma}$ has been taken to be independent of $t$.
Expressing these in terms of component fields,
\begin{eqnarray}
\frac{\partial N}{\partial t} + \mathrm{D}_a \left(F_N\right)^a &=& 0, 
\label{eq:Matter_31_components} \\
\frac{\partial \Pi_i}{\partial t} + \mathrm{D}_a  {\left(F_\Pi \right)^a}_i &=& 0, 
\label{eq:Momentum_31_components} \\
\frac{\partial E}{\partial t} + \mathrm{D}_a  \left(F_E \right)^a &=& 0, 
\label{eq:Energy_31_components} 
\end{eqnarray}
highlights that while the matter and energy densities $N$ and $E$ are scalar fields, and their fluxes $\bm{F}_N$ and $\bm{F}_E$ are vector fields, the momentum density $\bm{\Pi}$ is a linear form, and its flux $\bm{F}_{\bm{\Pi}}$ is a $(1,1)$ tensor.

For computations, these relations must be reduced to partial differential equations by further spelling out the 3-divergences on $\mathbb{S}_t$, yielding different expressions for different coordinate choices.
Consider three ways to obtain the resulting equations, taking note of what happens to the suggestive conservative form of the original geometric expressions in terms of divergences.

First, under present assumptions, because $\mathbb{S}_t$ is Euclidean (thanks to our focus here on flat $\mathbb{G}$ and $\mathbb{M}$) we could begin with rectangular coordinates $\left( x^{\bar\imath} \right)$, with respect to which the covariant derivative operator $\mathrm{D}_{\bar{\imath}}$ is simply a partial derivative $\partial / \partial x^{\bar{\imath}}$:
\begin{eqnarray}
\frac{\partial N}{\partial t} + \frac{\partial }{\partial x^{\bar{a}}} \left(F_N \right)^{\bar{a}} &=& 0, 
\label{eq:Matter_31_R} \\
\frac{\partial \Pi_{\bar{\imath}}}{\partial t} +  \frac{\partial }{\partial x^{\bar{a}}}  {\left(F_\Pi \right)^{\bar{a}}}_{\bar{\imath}} &=& 0, 
\label{eq:Momentum_31_R} \\
\frac{\partial E}{\partial t} + \frac{\partial }{\partial x^{\bar{a}}} \left(F_E \right)^{\bar{a}} &=& 0. 
\label{eq:Energy_31_R} 
\end{eqnarray}
This strict conservative form can be translated into numerical methods (such as finite-volume discretization) that naturally handle discontinuities and reproduce global conservation to numerical precision.
But under the coordinate transformation
\begin{equation}
\frac{\partial }{\partial x^{\bar{\imath}}} \rightarrow \frac{\partial u^a}{\partial x^{\bar{\imath}}} \,\frac{\partial }{\partial u^a}, 
\ \ \ \ F^{\bar{\imath}} \rightarrow \frac{\partial x^{\bar{\imath}}}{\partial u^a} \, F^a, 
\ \ \ \ {F^{\bar{\imath}}}_{\bar{\jmath}} \rightarrow \frac{\partial x^{\bar{\imath}}}{\partial u^a} \, 
\frac{\partial u^b}{\partial x^{\bar{\jmath}}}\, {F^a}_b, 
\end{equation}
additional terms arise from derivatives of the position-dependent Jacobian factors, and this apparently spoils the conservative form. 

Rather than proceed further with brute force, consider a second approach and make use of the rules of tensor analysis. 
In a coordinate basis, the covariant derivative of a tensor adds to the partial derivative an additional (sum of) ``connection'' terms for each tensor index. 
The gradient of a vector field yields the $(1,1)$ tensor field with components
\begin{equation}
\mathrm{D}_j F^i =  \frac{\partial F^i}{\partial u^j} + {\Gamma^i}_{a j} F^a,
\label{eq:GradientVector}
\end{equation}
in which the connection coefficients ${\Gamma^i}_{j k}$ (not tensor components in themselves!) are given by the Christoffel symbols
\begin{equation}
{\Gamma^i}_{j k} = \frac{1}{2} \gamma^{i a} 
	\left( \frac{\partial \gamma_{a k}}{\partial u^j} 
		+ \frac{\partial \gamma_{j a}}{\partial u^k}
		-  \frac{\partial \gamma_{j k}}{\partial u^a} \right)
\end{equation}
in terms of derivatives of the metric components. 
(The $(0,2)$ metric tensor $\bm{\gamma}$ being non-degenerate, the matrix $\left[ \gamma_{i j} \right]$ has matrix inverse $\left[ \gamma^{i j} \right]$ gathering the components of the $(2,0)$ inverse metric tensor $\overleftrightarrow{\bm{\gamma}}$.)
The divergence of a vector field is the contraction of Equation~(\ref{eq:GradientVector}).
It turns out that ${\Gamma^a}_{i a} = \partial \left( \ln \sqrt{\gamma} \right) / \partial u^i$, so that the divergence of a vector field can be expressed in the conservative form
\begin{equation}
\mathrm{D}_a F^a = \frac{1}{\sqrt{\gamma}} \, \frac{\partial }{\partial u^a} \left( \sqrt{\gamma} \, F^a \right).
\end{equation}
This is happily amenable to a structured grid in curvilinear coordinates, with a conservative finite-volume discretization of the divergence corresponding to the familiar definition encountered in elementary vector calculus:
\begin{equation}
 \frac{1}{\sqrt{\gamma}} \, \frac{\partial }{\partial u^a} \left( \sqrt{\gamma} \, F^a \right) \rightarrow
 \frac{1}{V_\leftrightarrow} \sum_a \left[ \left( A_a F^a \right)_{a \rightarrow} -  \left( A_a F^a \right)_{\leftarrow a} \right],
\end{equation}
in which $V_{\leftrightarrow}$ is a finite cell volume, and the cell face areas $A_i$ and flux components $F^i$ are evaluated on the outer ($i \rightarrow$) and inner ($\leftarrow i$) faces in each dimesnion $i$.
However, the momentum flux is a $(1,1)$ tensor field, and its covariant gradient has an additional term:
\begin{equation}
\mathrm{D}_k {F^i}_j =  \frac{\partial {F^i}_j}{\partial u^k} + {\Gamma^i}_{a k} {F^a}_j - {\Gamma^a}_{j k} {F^i}_a.
\label{eq:GradientTensor}
\end{equation}
Upon contraction to form the divergence $\mathrm{D}_a {F^a}_i$, the first connection term combines with the partial derivative as in the vector field case.
And when $F^{i j} = {F^i}_a \gamma^{a j}$ is symmetric---which is true for the momentum flux in fluid dynamics---there is some simplification in the second connection term, but it cannot be combined with the partial derivative.
(This simplification is a consequence of choosing to solve for the covariant momentum components $\Pi_i$ rather than the contravariant ones $\Pi^i$.)
The fluid equations take the form
\begin{eqnarray}
\frac{\partial N}{\partial t} +  \frac{1}{\sqrt{\gamma}} \, \frac{\partial }{\partial u^a} \left[ \sqrt{\gamma} \, \left( F_N \right)^a \right] &=& 0, 
\label{eq:Matter_31_C} \\
\frac{\partial \Pi_i}{\partial t} +   \frac{1}{\sqrt{\gamma}} \, \frac{\partial }{\partial u^a} \left[ \sqrt{\gamma} \, {\left( F_\Pi \right)^a}_i \right] &=& \frac{1}{2} \, \gamma^{b c} \, \frac{\partial \gamma_{c a}}{\partial u^i} \, {\left( F_\Pi \right)^a}_b, 
\label{eq:Momentum_31_C} \\
\frac{\partial E}{\partial t} +  \frac{1}{\sqrt{\gamma}} \, \frac{\partial }{\partial u^a} \left[ \sqrt{\gamma} \, \left( F_E \right)^a \right]  &=& 0. 
\label{eq:Energy_31_C} 
\end{eqnarray}
The matter and energy equations are in strict conservative form. 
And the momentum equations might also be said to be in conservative form in a looser sense: it is a balance equation, with a ``divergence'' on the left-hand side and without derivatives of the fluid variables in the source terms, so that finite volume discretization can still handle discontinuities.
But the terms on the right-hand side of Equation~(\ref{eq:Momentum_31_C}) constitute ``fictitious forces'': strictly speaking, curvilinear coordinates are non-inertial, and result in terms analogous to those that result from use of rotating or otherwise accelerated reference frames. 
Thus global conservation of curvilinear momentum components does not generally hold, and global conservation of rectangular momentum components will not be obtained to numerical precision.
(Note in passing that in spherical and cylindrical coordinates, the azimuthal component of momentum is the component of angular momentum along the azimuthal axis, the equation for which is in strict conservative form.
In cylindrical coordinates the equation for the linear momentum along the azimuthal axis is also in strict conservative form.) 

Use of a curvilinear coordinate mesh is sometimes indicated by the inherent geometry of a problem, even when three-dimensional phenomena preclude reduction in dimension because of an absence of axial or spherical symmetry.
In such a case, is there any way to avoid the fictitious forces on the right-hand side of Equation~(\ref{eq:Momentum_31_C})?

The purpose of this paper is to point out that the answer to this question is ``yes'', and to show it transparently, indeed almost instantly, with a third approach.
If one is willing or needs to work in three dimensions anyway, solving for \textit{rectangular} momentum components on a \textit{curvilinear} coordinate patch allows for exploitation of spherical or cylindrical geometry without the fictitious forces on the right-hand side of Equation~(\ref{eq:Momentum_31_C}).
This intuitively plausible result (mentioned without explanation or derivation in Ref.~\cite{Muller2020Hydrodynamics-o}) could of course be derived within the context of the first two approaches discussed in the preceding paragraphs.
But it becomes particularly obvious when we step back from brute force coordinate transformations, or rules for tensor analysis on components, and consider the differential geometry behind those approaches, treating tensors as geometric objects \cite{Gourgoulhon2013Special-Relativ,Gourgoulhon201231-Formalism-in}.

\section{Mixed basis for the momentum flux}

The point is that a tensor field is not just its component functions (even though these uniquely determine it). 
Components are merely expansion factors appearing when a tensor is expressed in terms of a basis---or, for fields, component functions appear when a tensor field is expressed in terms of a smoothly varying basis field.
The covariant derivative makes it possible to compare tensors at neighboring points of a manifold by taking account not only of the variation of component functions, but also the variation of the tensor basis field elements in terms of which the tensor field is expanded.

In particular, a $(1,1)$ tensor field is given by
\begin{equation}
\bm{F} = {F^a}_b \; \bm{e}_a \otimes \bm{e}^b
\end{equation}
in terms of a basis field $\left( \bm{e}_i \otimes \bm{e}^j \right)$ formed, via the tensor product, from some basis field $\left( \bm{e}_i \right)$ of vectors and some basis field $\left( \bm{e}^i \right)$ of linear forms.
The tensor field itself---a geometric object---is the same, regardless of whether one chooses (for example) the natural vector basis  $\left( \bm{\partial}_{x^{\bar\imath}} \right)$ and its dual linear form basis $\left( \bm{\mathrm{d}} x^{\bar\imath} \right)$ of rectangular coordinates, or the natural bases $\left( \bm{\partial}_{u^i} \right)$ and $\left( \bm{\mathrm{d}} u^i \right)$ of curvilinear coordinates: 
\begin{equation}
\bm{F} = {F^{\bar a}}_{\bar b} \; \bm{\partial}_{x^{\bar a}} \otimes \bm{\mathrm{d}} x^{\bar b}
	= {F^a}_b \; \bm{\partial}_{u^a} \otimes \bm{\mathrm{d}} u^ b.
\label{eq:TwoCoordinateSystems}
\end{equation}
We may be used to treating coordinate transformations as an all-or-nothing affair, for instance a choice between the two expressions in Equation~(\ref{eq:TwoCoordinateSystems}), to give equations in which all indices correspond to a particular choice of coordinates.

In fact, however, nothing prevents us from using the mixed curvilinear/rectangular basis
\begin{equation}
\bm{F} =  {F^a}_{\bar b} \; \bm{\partial}_{u^a} \otimes \bm{\mathrm{d}} x^{\bar b}
\label{eq:MixedBasis}
\end{equation}
on a coordinate patch with curvilinear coordinates $\left(u^i \right)$.
The covariant derivative obeys the Leibniz rule for derivatives of products, so that
\begin{equation}
\bm{\mathrm{D} F} =  \left( \left( \mathrm{D}_c {F^a}_{\bar b} \right) \; \bm{\partial}_{u^a} \otimes \bm{\mathrm{d}} x^{\bar b}
	+  {F^a}_{\bar b} \; \left(\mathrm{D}_c \, \bm{\partial}_{u^a} \right) \otimes \bm{\mathrm{d}} x^{\bar b}
	+  {F^a}_{\bar b} \; \bm{\partial}_{u^a} \otimes \left(\mathrm{D}_c \, \bm{\mathrm{d}} x^{\bar b}\right)
	\right) \otimes \bm{\mathrm{d}} u^c.
\label{eq:GradientTensorObject}
\end{equation}
Rules for tensor analysis notwithstanding, the component functions ${F^i}_{\bar\jmath}$ are actually scalar fields, for which the covariant derivative coincides with the partial derivative:
\begin{equation}
\mathrm{D}_k {F^i}_{\bar\jmath} = \partial_k {F^i}_{\bar\jmath}.
\end{equation}
Moreover the derivatives of the basis vector and linear form fields are precisely what give the connection coefficients in Equation~(\ref{eq:GradientTensor}):
\begin{equation}
\mathrm{D}_j \, \bm{\partial}_{u^i} = {\Gamma^a}_{ij}  \, \bm{\partial}_{u^a}, \ \ \ \  \mathrm{D}_j \, \bm{\mathrm{d}} u^i = - {\Gamma^i}_{aj}\, \bm{\mathrm{d}} u^a.
\end{equation}
But rectangular coordinate basis fields on Euclidean space are constant; any variation vanishes, including in particular
\begin{equation}
\mathrm{D}_j \, \bm{\mathrm{d}} x^{\bar\imath} = 0
\end{equation}
appearing in the third term of Equation~(\ref{eq:GradientTensorObject}).
Therefore
\begin{equation}
\bm{\mathrm{D} F} =  \left( \partial_c {F^a}_{\bar b} 
	+  {F^d}_{\bar b} {\Gamma^a}_{d c}
	\right) \, \bm{\partial}_{u^a} \otimes \bm{\mathrm{d}} x^{\bar b} \otimes \bm{\mathrm{d}} u^c.
\label{eq:GradientTensorObject_2}
\end{equation}
The divergence---the contraction on the first and third slots---takes the form
\begin{eqnarray}
\bm{\mathrm{D}} \cdot \bm{F} &=& \bm{\mathrm{D} F} \left( \bm{\mathrm{d}}u^a, \cdot, \bm{\partial}_{u^a}  \right) \nonumber \\
	&=& \left( \partial_a {F^a}_{\bar b} +  {\Gamma^a}_{c a} {F^c}_{\bar b} \right) \, \bm{\mathrm{d}} x^{\bar b},
\label{eq:DivergenceConservative}
\end{eqnarray}
since
\begin{eqnarray}
\left( \bm{\partial}_{u^i} \otimes \bm{\mathrm{d}} x^{\bar\jmath} \otimes \bm{\mathrm{d}} u^k \right) \left( \bm{\mathrm{d}}u^a, \cdot, \bm{\partial}_{u^a}  \right)  
 &=& \left( \bm{\partial}_{u^i} \left(  \bm{\mathrm{d}}u^a \right) \right) \left( \bm{\mathrm{d}} u^k \left(  \bm{\partial}_{u^a} \right) \right) \,  \bm{\mathrm{d}} x^{\bar\jmath} \nonumber \\
 	&=& \delta^a_i \delta^k_a \, \bm{\mathrm{d}} x^{\bar\jmath} \nonumber \\
	&=&\delta^k_i \, \bm{\mathrm{d}} x^{\bar\jmath}.
\end{eqnarray}
Recall also that ${\Gamma^a}_{i a} = \partial \left( \ln \sqrt{\gamma} \right) / \partial u^i$, as noted previously.

Thus Equation~(\ref{eq:DivergenceConservative}) shows that the fictitious forces on the right-hand side of Equation~(\ref{eq:Momentum_31_C}), which stem from the connection term on the covariant index of a $(1,1)$ tensor $\bm{F}$, vanish when $\bm{F}$ is expanded in terms of the mixed basis of Equation~(\ref{eq:MixedBasis}).
Then, on flat spacetimes $\mathbb{G}$ and $\mathbb{M}$ and in the absence of external forces and heating/cooling, all the fluid equations
\begin{eqnarray}
\frac{\partial N}{\partial t} +  \frac{1}{\sqrt{\gamma}} \, \frac{\partial }{\partial u^a} \left[ \sqrt{\gamma} \, \left( F_N \right)^a \right] &=& 0, 
\label{eq:Matter_31_M} \\
\frac{\partial \Pi_{\bar\imath}}{\partial t} +   \frac{1}{\sqrt{\gamma}} \, \frac{\partial }{\partial u^a} \left[ \sqrt{\gamma} \, {\left( F_\Pi \right)^a}_{\bar\imath} \right] &=& 0, 
\label{eq:Momentum_31_M} \\
\frac{\partial E}{\partial t} +  \frac{1}{\sqrt{\gamma}} \, \frac{\partial }{\partial u^a} \left[ \sqrt{\gamma} \, \left( F_E \right)^a \right]  &=& 0. 
\label{eq:Energy_31_M} 
\end{eqnarray}
are in strict conservative form.
Again, it is the rectangular momentum component fields $\Pi_{\bar\imath}$ being solved for, from the expansion $\bm{\Pi} = \Pi_{\bar a} \, \bm{\mathrm{d}} x^{\bar a}$, rather than the curvilinear component field from the alternative expansion $\bm{\Pi} = \Pi_a \, \bm{\mathrm{d}} u^a$.

\section{Fluid dynamics equations in curvilinear and mixed bases}

By way of concrete example, consider the fluid dynamics equations on $\mathbb{M}$ and $\mathbb{G}$, specializing to a perfect fluid, in curvilinear coordinates general enough to encompass the rectangular, cylindrical, and spherical cases, before and after eliminating the fictitious forces associated with these coordinate choices.
Before arriving at partial differential equations in particular coordinates it is necessary to give a $3+1$ decomposition of the spacetime tensor Equations~(\ref{eq:MatterDivergence_4}) and (\ref{eq:EnergyMomentumDivergence_4}) satisfied by the matter and relative energy-momentum 4-fluxes, taking account also of the necessity to find expressions in terms of quantities measured by a comoving observer in order to apply constitutive equations (in this case, an equation of state).
For further background, and on allowance for heat flow and viscosity, see Refs.~\cite{Cardall2019Minkowski-and-G,Cardall2020Combining-3-Mom} (beware however some notational changes).

\subsection{$3+1$ decomposition of the matter and relative energy-momentum fluxes}

Recall the foliation of affine spacetimes $\mathbb{M}$ and $\mathbb{G}$ introduced in Section~\ref{sec:Introduction}.
Fiducial (and here, inertial) observers have straight worldlines with 4-velocity vector field $\bm{w} = \bm{\partial}_t$ threading spacelike slices $\left(\mathbb{S}_t \right)$, affine hyperplanes, the level surfaces of global time coordinate $t$. 
Introduce now the the linear form $\bm{t} = \bm{\nabla} t = \bm{\mathrm{d}}t$ associated with the $\left(\mathbb{S}_t \right)$.
Because $\bm{\mathrm{d}}t$ is dual to $\bm{\partial}_t$,
\begin{equation}
\bm{w} \cdot \bm{t} = \bm{t} \cdot \bm{w} = 1, 
\end{equation}
while for a vector $\bm{a}$ tangent to $\mathbb{S}_t$,
\begin{equation}
\bm{a} \cdot \bm{t} = \bm{t} \cdot \bm{a} = 0. 
\end{equation}
(Note that in this paper the dot, $\cdot$, only denotes contraction and never an inner product.)
We have already introduced $\bm{\gamma}$ as the 3-metric on $\mathbb{S}_t$.
It is consistent to use the same notation to denote the related projection tensor from $\mathbb{M}$ or $\mathbb{G}$ to $\mathbb{S}_t$. 
In particular we need the related projection tensor
\begin{equation}
\overleftarrow{\bm{\gamma}} = \bm{\delta} - \bm{w} \otimes \bm{t}
\end{equation}
on $\mathbb{S}_t$, where $\bm{\delta}$ is the $\left(1,1\right)$ identity tensor on $\mathbb{M}$ or $\mathbb{G}$, for which 
\begin{equation}
\overleftarrow{\bm{\gamma}} \cdot \bm{w} = 0, \ \ \ \ \bm{t} \cdot \overleftarrow{\bm{\gamma}} = 0,
\end{equation}
and 
\begin{equation}
\overleftarrow{\bm{\gamma}} \cdot \bm{a} = \bm{a}, \ \ \ \ \underline{\bm{a}} \cdot \overleftarrow{\bm{\gamma}}  = \underline{\bm{a}} 
\end{equation}
for $\bm{a}$ tangent to $\mathbb{S}_t$ and the linear form
\begin{equation}
 \underline{\bm{a}} = \bm{\gamma} \cdot \bm{a} = \bm{a} \cdot \bm{\gamma},
\end{equation}
the metric dual of $\bm{a}$ on $\mathbb{S}_t$.

The conservative formulations we desire follow from decomposing vector fields on $\mathbb{M}$ or $\mathbb{G}$ into pieces parallel to $\bm{w}$ and tangent to $\mathbb{S}_t$.
Under the geometry described above, a vector field $\bm{A}$ so decomposed,
\begin{equation}
\bm{A} = A \bm{w} + \bm{a} = A \bm{w} + \overleftarrow{\bm{\gamma}} \cdot \bm{a}, 
\end{equation}
has 4-divergence
\begin{eqnarray}
\bm{\nabla} \cdot \bm{A} &=& \left( \bm{w} \cdot \bm{\nabla} \right) A 
+ \left( \overleftarrow{\bm{\gamma}} \cdot \bm{\nabla} \right) \cdot a \nonumber \\
	&=& \frac{\partial A}{\partial t} + \bm{\mathrm{D}} \cdot \bm{a}.
\label{eq:GeneralDivergence}
\end{eqnarray}
We so decompose the matter flux $\bm{N}$ and relative energy-momentum flux $\bm{S}$.

Let the flow of matter---taken to be baryon number---define a field of comoving observers.
That is, 
\begin{equation}
\bm{N} = n \bm{U} \ \ \ \ \mathrm{on\ }\mathbb{M}\mathrm{\ and\ }\mathbb{G},
\end{equation}
where $n$ is the baryon number density measured by the comoving observers, and $\bm{U}$ is their 4-velocity field.
From the perspective of the fiducial observers,
\begin{equation}
\bm{U} = \left\{ \begin{array}{rl}
	\Lambda_{\bm{v}} \left( \bm{w} + \bm{v} \right) &  \mathrm{on\ }\mathbb{M}, \\ 
	 \bm{w} + \bm{v} &  \mathrm{on\ }\mathbb{G},
	\end{array} \right.
\end{equation}
where $\bm{v}$ is tangent to $\mathbb{S}_t$ and is the 3-velocity of the fluid measured by the fiducial observers.
On $\mathbb{M}$, the Lorentz factor $\Lambda_{\bm{v}} = \sqrt{1 - \bm{\gamma}(\bm{v},\bm{v}) / c^2}$ follows from the normalization $\bm{g}\left( \bm{U}, \bm{U} \right) = -c^2$ given by the Lorentz metric $\bm{g}$. 
Comparing with the decomposition 
\begin{equation}
\bm{N} = N \bm{w} + \bm{F}_N  \ \ \ \ \mathrm{on\ }\mathbb{M}\mathrm{\ and\ }\mathbb{G},
\end{equation}
we find 
\begin{equation}
N = \left\{ \begin{array}{rl} 
	\Lambda_{\bm{v}} n &  \mathrm{on\ }\mathbb{M}, \\ 
	n & \mathrm{on\ }\mathbb{G}, 
	\end{array} \right.
\label{eq:FiducialDensity}
\end{equation}
and
\begin{equation}
\bm{F}_N = N \bm{v}  \ \ \ \ \mathrm{on\ }\mathbb{M}\mathrm{\ and\ }\mathbb{G}.
\end{equation}
Therefore, in accord with Equation~(\ref{eq:GeneralDivergence}), $\bm{\nabla} \cdot \bm{N} = 0$ becomes
\begin{equation}
\frac{\partial N}{\partial t} + \bm{\mathrm{D}} \cdot \left(  N  \bm{v} \right) = 0  \ \ \ \ \mathrm{on\ }\mathbb{M}\mathrm{\ and\ }\mathbb{G}, 
\label{eq:Matter_31_Flux}
\end{equation}
with however the differing baryon densities $N$ measured by fiducial observers on $\mathbb{M}$ and $\mathbb{G}$ as given by Equation~(\ref{eq:FiducialDensity}).

The flow of relative energy-momentum calls for some explanation \cite{Cardall2020Combining-3-Mom}.
It is a $(1,1)$ tensor
\begin{equation}
\bm{S} = n\bm{U} \otimes \bm{P} - \bm{\Sigma}  \ \ \ \ \mathrm{on\ }\mathbb{M}\mathrm{\ and\ }\mathbb{G}
\label{eq:RelativeEnergyMomentumFlux}
\end{equation}
expressed in terms of two primary pieces.

The first term in Equation~(\ref{eq:RelativeEnergyMomentumFlux}) is the relative 4-momentum per baryon $\bm{P}$, a linear form, carried by the baryon flux $n \bm{U}$.
To begin to understand $\bm{P}$, consider first the existence on both $\mathbb{M}$ and $\mathbb{G}$  of an inertia flux vector $\bm{I} = m \bm{U}$, where $m$ is the mass per baryon.
The inertia flux has timelike component $\Lambda_{\bm{v}} m$ on $\mathbb{M}$ and $m$ on $\mathbb{G}$, the inertia per baryon measured by a fiducial observer.
The metric dual of $\bm{I}$ on $\mathbb{M}$ with respect to $\bm{g}$, the linear form $\bm{g} \cdot \bm{I}$, is the traditional relativistic 4-momentum per baryon, with timelike component $-\Lambda_{\bm{v}} m c^2$, the (negative of the) mass-energy per baryon as measured by a fiducial observer.
But this timelike component becomes meaningless as $c \rightarrow \infty$: unlike the inertia flux, the traditional relativistic 4-momentum makes no sense on $\mathbb{G}$, which allows no mass-energy equivalence.
What does make sense on both $\mathbb{M}$ and $\mathbb{G}$ is to define a ``relative'' 4-momentum per baryon $\bm{P}$, a linear form whose timelike component is the kinetic energy measured by the fiducial observers.
On $\mathbb{M}$ it can be expressed as $\bm{g} \cdot m \left( \bm{U} - \bm{w} \right)$, yielding
\begin{equation}
\bm{P} = \left\{ \begin{array}{rl}
- \left( \Lambda_{\bm{v}} - 1 \right) m c^2 \, \bm{t} + \Lambda_{\bm{v}} m \, \underline{\bm{v}}  & \mathrm{on\ }\mathbb{M}, \\
-\frac{1}{2} m \bm{\gamma}\left( \bm{v}, \bm{v} \right) \, \bm{t} + m \, \underline{\bm{v}} & \mathrm{on\ }\mathbb{G},
\end{array}\right.
\label{eq:RelativeFourMomentum}
\end{equation}
with the definition on $\mathbb{G}$ taken to be the $c \rightarrow \infty$ limit of the expression on $\mathbb{M}$.
Recall that $\underline{\bm{v}} = \bm{\gamma} \cdot \bm{v}$ is a linear form on $\mathbb{S}_t$, the metric dual of the 3-velocity $\bm{v}$, which is tangent to $\mathbb{S}_t$.

The second term in Equation~(\ref{eq:RelativeEnergyMomentumFlux}), the (negative of the) 4-stress $\bm{\Sigma}$, a $(1,1)$ tensor, arises from considering each fluid element to be an infinitesimal thermodynamic system in its own right (this tensor can also be arrived at using a variational approach \cite{Duval1978Dynamics-of-con}).
For physical interpretation it is once again convenient to begin with a raised index, considering the $(2,0)$ tensor $\overrightarrow{\bm{\Sigma}} = \bm{\Sigma} \cdot \overleftrightarrow{\bm{g}}$ on $\mathbb{M}$ and the (negative of) inertia-momentum flux it represents.
For a perfect fluid, 
\begin{eqnarray}
- \overrightarrow{\bm{\Sigma}} &=& n\bm{U} \otimes \frac{e \bm{U}}{c^2 n} + p \, \overleftrightarrow{\bm{\gamma}} \nonumber \\
&=& n\bm{U} \otimes \frac{e \bm{U}}{c^2 n} + p \left( \overleftrightarrow{\bm{g}}+\frac{1}{c^2} \bm{U} \otimes \bm{U} \right) 
\ \ \ \ \mathrm{on\ }\mathbb{M}.
\end{eqnarray}
The first term is the flux of inertia per baryon due to internal energy density $e$, measured by comoving observers (and with no heat flux out of fluid elements), carried by the baryon flux $n \bm{U}$. 
In the second term, the projection tensor orthogonal to $\bm{U}$ enforces the spatial stress as seen by comoving observers to be isotropic and given by a pressure $p$.
Lowering the second index gives
\begin{equation}
- \bm{\Sigma} = \left\{ \begin{array}{rl}
- \bm{U} \otimes \Lambda_{\bm{v}} \left( e +p \right) \left( \bm{t} - \frac{1}{c^2}\, \underline{\bm{v}} \right) + p \, \bm{\delta} & \mathrm{on\ }\mathbb{M}, \\ 
- \bm{U} \otimes \left( e +p \right) \, \bm{t}  + p \, \bm{\delta} & \mathrm{on\ }\mathbb{G},  
\end{array}\right.
\label{eq:FourStress}
\end{equation}
with the definition on $\mathbb{G}$ following once again as the $c\rightarrow \infty$ limit of the expression on $\mathbb{M}$.

Note that $\bm{S}$ must be a $(1,1)$ tensor in order combine energy and momentum on $\mathbb{G}$ (even though it can only be kinetic plus internal energy---that is, macroscopic plus microscopic kinetic energy) \cite{Cardall2020Combining-3-Mom}.
A contravariant index is needed to have a flow on spacetime, and a divergence of that flow.
But a covariant index is also needed, in order for energy in a timelike component to survive the $c\rightarrow \infty$ limit. 
This seems to be a reflection of absolute inertia corresponding to absolute time on $\mathbb{G}$.
In any case, that 4-momentum be consistently regarded as naturally a linear form, while 4-velocity is a vector, is consistent with the Hamiltonian perspective of (4-)momentum being conjugate to (4-)position.

Putting Equations~(\ref{eq:RelativeFourMomentum}) and (\ref{eq:FourStress}) in Equation~(\ref{eq:RelativeEnergyMomentumFlux}), the relative energy-momentum flux can be expressed as
\begin{equation}
\bm{S} = \left(\bm{w} + \bm{v} \right) \otimes \left( -\left( E + p \right) \bm{t} + \bm{\Pi} \right) + p \, \bm{\delta}  \ \ \ \ \mathrm{on\ }\mathbb{M}\mathrm{\ and\ }\mathbb{G},
\label{eq:RelativeEnergyMomentumFlux_2}
\end{equation}
where
\begin{equation}
\bm{\Pi} = \left\{ \begin{array}{rl}
\Lambda_{\bm{v}} \left( m n + \frac{1}{c^2}\left(e+p\right) \right) \, \underline{\bm{v}} & \mathrm{on\ }\mathbb{M}, \\ 
m n \, \underline{\bm{v}} & \mathrm{on\ }\mathbb{G},
\end{array}\right.
\label{eq:MomentumDensity_F}
\end{equation}
and
\begin{equation}
E = \left\{ \begin{array}{rl}
\Lambda_{\bm{v}} \left( \left( \Lambda_{\bm{v}} - 1 \right) m n c^2 + \Lambda_{\bm{v}} \left( e +p \right) \right) - p & \mathrm{on\ }\mathbb{M}, \\ 
\frac{1}{2} m \bm{\gamma}\left( \bm{v}, \bm{v} \right) + e & \mathrm{on\ }\mathbb{G}.
\end{array}\right.
\label{eq:EnergyDensity_F}
\end{equation}
In a moment it will become clear that the linear form $\bm{\Pi}$ is the momentum density of the fluid measured by fiducial observers, and that the scalar field $E$ is the relative energy density of the fluid measured by fiducial observers.

Because $\bm{S}$ is a rank-2 tensor, its $3+1$ decomposition must proceed on both indices.
Begin with the second (covariant) index, using $\bm{S}$ as given in Equation~(\ref{eq:RelativeEnergyMomentumFlux_2}).
Isolate the momentum flux $\bm{M}$, also a $(1,1)$ tensor field, as the spatial projection of $\bm{S}$ according to fiducial observers:
\begin{eqnarray}
\bm{M} &=& \bm{S} \cdot \overleftarrow{\bm \gamma} \nonumber \\
	&=& \left(\bm{w} + \bm{v} \right) \otimes \bm{\Pi} + p \, \overleftarrow{\bm \gamma} \nonumber \\
	&=&  \bm{w} \otimes \bm{\Pi} + \left( \bm{v}  \otimes \bm{\Pi}  + p \, \overleftarrow{\bm \gamma} \right) \ \ \ \ \mathrm{on\ }\mathbb{M}\mathrm{\ and\ }\mathbb{G},
\label{eq:MomentumFlux}
\end{eqnarray}
where the terms in parentheses in the last line are the momentum 3-flux on $\mathbb{S}_t$.
Isolate the relative energy flux $\bm{E}$, a vector field, as the projection of $\bm{S}$ along $\bm{w}$:
\begin{eqnarray}
\bm{E} &=& -\bm{S} \cdot \bm{w} \nonumber \\
	&=& \left( E + p \right) \left(\bm{w} + \bm{v} \right) - p \, \bm{w} \nonumber \\
	&=& E\, \bm{w} +  \left( E + p \right)\, \bm{v} \ \ \ \  \ \ \ \ \ \ \ \ \ \ \ \ \mathrm{on\ }\mathbb{M}\mathrm{\ and\ }\mathbb{G}.
\label{eq:RelativeEnergyFlux}
\end{eqnarray}
For our inertial observers on flat spacetimes $\mathbb{M}$ and $\mathbb{G}$, the covariant derivatives of $\bm{w}$ and  $\overleftarrow{\bm \gamma}$ vanish. 
Therefore $\bm{\nabla} \cdot \bm{S} = 0$ breaks up into $\bm{\nabla} \cdot \bm{M} = 0$ and $\bm{\nabla} \cdot \bm{E} = 0$; and according to Equation~(\ref{eq:GeneralDivergence}), these give
\begin{eqnarray}
\frac{\partial \bm{\Pi}}{\partial t} + \bm{\mathrm{D}} \cdot \left( \bm{v}  \otimes \bm{\Pi}  + p \, \overleftarrow{\bm \gamma}  \right) &=& 0  \ \ \ \ \mathrm{on\ }\mathbb{M}\mathrm{\ and\ }\mathbb{G}, 
\label{eq:Momentum_31_Flux} \\
\frac{\partial E}{\partial t} + \bm{\mathrm{D}} \cdot \left(  \left( E + p \right) \, \bm{v} \right) &=& 0  \ \ \ \ \mathrm{on\ }\mathbb{M}\mathrm{\ and\ }\mathbb{G}, 
\label{eq:RelativeEnergy_31_Flux}
\end{eqnarray}
with however differing momentum and relative energy densities $\bm{\Pi}$ and $E$ measured by fiducial observers on $\mathbb{M}$ and $\mathbb{G}$, as given by Equations~(\ref{eq:MomentumDensity_F}) and (\ref{eq:EnergyDensity_F}).

\subsection{Curvilinear coordinates and reduction to partial differential equations}

Now Equations~(\ref{eq:Matter_31_Flux}), (\ref{eq:Momentum_31_Flux}), and (\ref{eq:RelativeEnergy_31_Flux}) must be reduced to partial differential equations through a specification of coordinates on $\mathbb{S}_t$.
Rectangular coordinates $(x, y, z)$, cylindrical coordinates $(\varrho, z, \phi)$, and spherical coordinates $(r, \theta, \phi)$ can all be regarded as coordinates $(u, v, w)$ with metric components gathered by the matrix
\begin{equation}
\left[ \gamma_{i j} \right] = 
 \begin{bmatrix}
 1 & 0 & 0 \\
 0 & f(u)^2 & 0 \\
 0 & 0 & g ( u )^2 \, h ( v )^2
 \end{bmatrix},
 \end{equation}
with $\sqrt{\gamma} = f g h$ (see Table~\ref{tab:CurvilinearMetric}; the curvilinear coordinate $v$ is not to be confused with the coordinates $v^i$ of the 3-velocity $\bm{v}$).
Consulting Equations~(\ref{eq:Matter_31_C})-(\ref{eq:Energy_31_C}) and (\ref{eq:Matter_31_Flux}), (\ref{eq:Momentum_31_Flux})-(\ref{eq:RelativeEnergy_31_Flux}), the fluid dynamics equations read
 \begin{eqnarray}
\frac{\partial N}{\partial t} +  \frac{1}{f g h} \, \frac{\partial }{\partial u^a} \left[ f g h \, \left( n v^a \right) \right] &=& 0, 
\label{eq:Matter_Curvilinear} \\
\frac{\partial \Pi_1}{\partial t} +   \frac{1}{f g h} \, \frac{\partial }{\partial u^a} \left[ f g h \, \left( v^a \Pi_1 + p \,  {\delta^a}_1 \right) \right] &=& 
	\frac{1}{f} \frac{\mathrm{d} f}{\mathrm{d} u} \left( v^2 \Pi_2 + p \right)
	+ \frac{1}{g} \frac{\mathrm{d} g}{\mathrm{d} u} \left( v^3 \Pi_3 + p \right),
\label{eq:Momentum_Curvilinear_1} \\
\frac{\partial \Pi_2}{\partial t} +   \frac{1}{f g h} \, \frac{\partial }{\partial u^a} \left[ f g h \, \left( v^a \Pi_2 + p \,  {\delta^a}_2 \right) \right] &=& 
	\frac{1}{h} \frac{\mathrm{d} h}{\mathrm{d} v} \left( v^3 \Pi_3 + p \right),
\label{eq:Momentum_Curvilinear_2} \\
\frac{\partial \Pi_3}{\partial t} +   \frac{1}{f g h} \, \frac{\partial }{\partial u^a} \left[ f g h \, \left( v^a \Pi_3 + p \,  {\delta^a}_3 \right) \right] &=& 0,
\label{eq:Momentum_Curvilinear_3} \\
\frac{\partial E}{\partial t} +  \frac{1}{f g h} \, \frac{\partial }{\partial u^a} \left[ f g h \, \left( E + p \right) v^a \right]  &=& 0. 
\label{eq:Energy_Curvilinear} 
\end{eqnarray}
But in a mixed basis for the momentum equations, in which we solve for the rectangular components of momentum density, we have from Equations~(\ref{eq:Matter_31_M})-(\ref{eq:Energy_31_M})
\begin{eqnarray}
\frac{\partial N}{\partial t} +  \frac{1}{f g h} \, \frac{\partial }{\partial u^a} \left[ f g h \, \left( n v^a \right) \right] &=& 0, 
\label{eq:Matter_Mixed} \\
\frac{\partial \Pi_{\bar\imath}}{\partial t} +   \frac{1}{f g h} \, \frac{\partial }{\partial u^a} \left[ f g h \, {\left(  v^a \Pi_{\bar\imath} + p \, \frac{\partial u^a}{\partial x^{\bar\imath}} \right)} \right] &=& 0, 
\label{eq:Momentum_Mixed} \\
\frac{\partial E}{\partial t} +  \frac{1}{f g h} \, \frac{\partial }{\partial u^a} \left[ f g h \, \left( E + p \right) v^a \right]  &=& 0. 
\label{eq:Energy_Mixed} 
\end{eqnarray}
Fictitious forces no longer appear on the right-hand side.
But the momentum flux is now more complicated: instead of a Kronecker $\delta$ multiplying the pressure in the momentum flux, the inverse Jacobian $\partial u^a / \partial x^{\bar\imath}$ appears; and this inverse Jacobian must also be used in obtaining curvilinear components $v^i$ from the rectangular momentum density components $\Pi_{\bar\imath}$ after the latter have been updated.
For spherical coordinates $(r,\theta,\phi)$ given by $(x,y,z) = (r \sin\theta \cos\phi, r \sin\theta \sin\phi, r \cos\phi)$, the inverse Jacobian is
\begin{equation}
\left[ \frac{\partial u^j}{\partial x^{\bar\imath}} \right] 
	= \frac{1}{r} \begin{bmatrix} r \sin\theta \cos\phi & r \sin\theta \sin\phi & r \cos\theta \\
	\cos\theta \cos\phi & \cos\theta \sin\phi & - \sin\theta \\
	-\sin\phi / \sin\theta & \cos\phi / \cos\theta & 0
	\end{bmatrix}.
\end{equation}
For cylindrical coordinates $(\varrho,z,\phi)$ given by $(x,y,z) = (\varrho \cos\phi, \varrho  \sin\phi, z)$, the inverse Jacobian is
\begin{equation}
\left[ \frac{\partial u^j}{\partial x^{\bar\imath}} \right] 
	= \frac{1}{\varrho} \begin{bmatrix} \varrho \cos\phi & \varrho \sin\phi & 0 \\
	0 & 0 & 1 \\
	- \sin\phi &  \cos\phi & 0
	\end{bmatrix}.
\end{equation}
The dependence of the flux components on $\theta$ and $\phi$ means that computations must be performed in three dimensions: reduction of dimension under conditions of axial or spherical symmetry is not an option when solving for the rectangular components of momentum.

\begin{table}
\caption{Coordinates, metric functions, and metric derivatives for common curvilinear coordinate systems.}
\centering
\begin{tabular}{cccccccccc}
\toprule
\textbf{System}	& $u$ & $v$ &	$w$ & $f$ & $g$ & $h$ & $\frac{1}{f} \frac{\mathrm{d} f}{\mathrm{d} u}$ & $\frac{1}{g} \frac{\mathrm{d} g}{\mathrm{d} u}$ & $\frac{1}{h} \frac{\mathrm{d} h}{\mathrm{d} v}$ \\
\midrule
Rectangular & $x$ & $y$ & $z$ & $1$ & $1$ & $1$ & $0$ & $0$ & $0$ \\
Cylindrical & $\varrho$ & $z$ & $\phi$ & $1$ & $\varrho$ & $1$ & $0$ & $\frac{1}{\varrho}$ & 0 \\
Spherical & $r$ & $\theta$ & $\phi$ & $r$ & $r$ & $\sin\theta$ & $\frac{1}{r}$ & $\frac{1}{r}$ & $\frac{\cos\theta}{\sin\theta}$ \\
\bottomrule
\end{tabular}
\label{tab:CurvilinearMetric}
\end{table}

\section{Conclusion}

Some fluid dynamics problems are couched most naturally in terms of spherical or cylindrical coordinates, even when three-dimensional phenomena are of interest in a generally spherical or cylindrical geometric scenario.
When one needs or is willing to work in three dimensions anyway, the fictitious forces arising from curvilinear coordinates can be eliminated by solving for rectangular momentum components on a curvilinear mesh.
This allows a strictly conservative form of the fluid dynamics equations to be restored on flat spacetime and in the absence of external forces and heating/cooling.
However, angular dependence is thereby introduced into the momentum flux: the advected rectangular momentum components do not exhibit spherical or cylindrical symmetry, and the stress is transformed by a Jacobian matrix that explicitly introduces angular dependence.
(Thus the necessity to work in three dimensions if this option is to be exercised.)
But fictitious force terms that grow large near coordinate singularities are absorbed into the flux, where they are partially ameliorated when multiplied by the metric determinant appearing in the divergence.

When spherical- or cylindrical-style coordinates are used in the curved spacetime of general relativity, this approach can still eliminate the source terms associated with those coordinates that may cause trouble near coordinate singularities, even as more physically relevant source terms representing gravitation remain. 

The possibility of eliminating fictitious forces arising from curvilinear coordinates is made transparent by an approach and notation that treat tensors as geometric objects.
The thoroughly geometric approach modeled here will also improve the transparency of formulations involving curved spacetime or otherwise accelerated, i.e. Lagrange or arbitrary-Euler-Lagrange, observers.
(The difference from the present focus on inertial fiducial observers in flat spacetime will be that the 4-velocity and spatial projection tensor fields associated with accelerated observers will not have vanishing covariant derivative.) 
This geometric approach also highlights a greater unity between relativistic and non-relativistic fluid dynamics than is usually appreciated. 

\funding{This work was supported by the U.S. Department of Energy, Office of Science, Office of Nuclear Physics under contract number DE-AC05-00OR22725.}

\conflictsofinterest{The author declares no conflict of interest. The funders had no role in the design of the study; in the collection, analyses, or interpretation of data; in the writing of the manuscript, or in the decision to publish the results.}

\end{paracol}
\reftitle{References}

\end{document}